\documentclass[8.5pt,twoside,twocolumn]{article}
\usepackage[super,sort&compress,comma]{natbib}
\usepackage{mhchem}
\usepackage{times,mathptmx}
\usepackage[T1]{fontenc} 
\usepackage{sectsty}
\usepackage{balance}
\usepackage{color}
\usepackage{amssymb}

\usepackage{graphicx} 
\usepackage{lastpage}
\usepackage[format=plain,justification=raggedright,singlelinecheck=false,font=small,labelfont=bf,labelsep=space]{caption}
\usepackage{fancyhdr}
\begin{document}

\thispagestyle{plain}
\fancypagestyle{plain}{
\fancyhead[L]{\includegraphics[height=8pt]{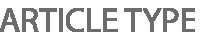}}
\fancyhead[C]{\hspace{-1cm}\includegraphics[height=20pt]{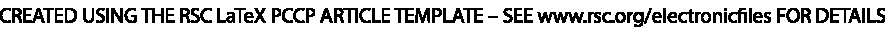}}
\fancyhead[R]{\includegraphics[height=10pt]{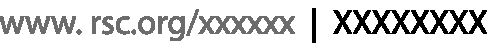}\vspace{-0.2cm}}
\renewcommand{\headrulewidth}{1pt}}
\renewcommand{\thefootnote}{\fnsymbol{footnote}}
\renewcommand\footnoterule{\vspace*{1pt}%
\hrule width 3.4in height 0.4pt \vspace*{5pt}}
\setcounter{secnumdepth}{5}

\makeatletter
\def\subsubsection{\@startsection{subsubsection}{3}{10pt}{-1.25ex plus -1ex minus -.1ex}{0ex plus 0ex}{\normalsize\bf}}
\def\paragraph{\@startsection{paragraph}{4}{10pt}{-1.25ex plus -1ex minus -.1ex}{0ex plus 0ex}{\normalsize\textit}}
\renewcommand\@biblabel[1]{#1}
\renewcommand\@makefntext[1]%
{\noindent\makebox[0pt][r]{\@thefnmark\,}#1}
\makeatother
\renewcommand{\figurename}{\small{Fig.}~}
\sectionfont{\large}
\subsectionfont{\normalsize}

\fancyfoot{}
\fancyfoot[LO,RE]{\vspace{-7pt}\includegraphics[height=9pt]{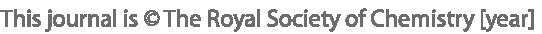}}
\fancyfoot[CO]{\vspace{-7.2pt}\hspace{12.2cm}\includegraphics{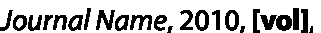}}
\fancyfoot[CE]{\vspace{-7.5pt}\hspace{-13.5cm}\includegraphics{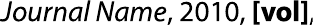}}
\fancyfoot[RO]{\footnotesize{\sffamily{1--\pageref{LastPage} ~\textbar  \hspace{2pt}\thepage}}}
\fancyfoot[LE]{\footnotesize{\sffamily{\thepage~\textbar\hspace{3.45cm} 1--\pageref{LastPage}}}}
\fancyhead{}
\renewcommand{\headrulewidth}{1pt}
\renewcommand{\footrulewidth}{1pt}
\setlength{\arrayrulewidth}{1pt}
\setlength{\columnsep}{6.5mm}
\setlength\bibsep{1pt}

\twocolumn[
  \begin{@twocolumnfalse}
\noindent\LARGE{\textbf{Compositional dependence of anomalous thermal expansion in perovskite-like ABX$_3$ formates$^\dag$}}
\vspace{0.6cm}

\noindent\large{\textbf{Ines E.\ Collings,\textit{$^{a,b}$} Joshua A.\ Hill,\textit{$^{a}$} Andrew B.\ Cairns,\textit{$^{a}$} Richard I.\ Cooper,\textit{$^{a}$} Amber L.\ Thompson,\textit{$^{a}$} Julia E.\ Parker,\textit{$^{c}$} Chiu C.\ Tang,\textit{$^{c}$}   and Andrew L.\ Goodwin$^{\ast}$\textit{$^{a}$} }}\vspace{0.5cm}

\noindent\textit{\small{\textbf{Received Xth XXXXXXXXXX 20XX, Accepted Xth XXXXXXXXX 20XX\newline
First published on the web Xth XXXXXXXXXX 200X}}}

\noindent \textbf{\small{DOI: 10.1039/b000000x}}
\vspace{0.6cm}

\noindent \normalsize{The compositional dependence of thermal expansion behaviour in 19 different perovskite-like metal--organic frameworks (MOFs) of composition [A$^{\textrm{I}}$][M$^{\textrm{II}}$(HCOO)$_3$] (A = alkylammonium cation; M = octahedrally-coordinated divalent metal) is studied using variable-temperature X-ray powder diffraction measurements. While all systems show essentially the same type of thermomechanical response---irrespective of their particular structural details---the magnitude of this response is shown to be a function of A$^{\textrm{I}}$ and M$^{\textrm{II}}$ cation radii, as well as the molecular anisotropy of  A$^{\textrm{I}}$. Flexibility is maximised for large M$^{\textrm{II}}$ and small A$^{\textrm{I}}$, while the shape of A$^{\textrm{I}}$ has implications for the direction of framework hingeing.}
\vspace{0.5cm}
 \end{@twocolumnfalse}
  ]

\footnotetext{\dag~Electronic Supplementary Information (ESI) available: Details of Gua-Cd single-crystal X-ray diffraction, Rietveld fits, variable-temperature lattice parameters, XBU $r$ and $\theta$ equations, and A-site cation size details.}
\footnotetext{\textit{$^{a}$~Department of Chemistry, University of Oxford, Inorganic Chemistry Laboratory, South Parks Road, Oxford OX1 3QR, U.K. Fax: +44 1865 274690; Tel: +44 1865 272137; E-mail: andrew.goodwin@chem.ox.ac.uk.}}
\footnotetext{\textit{$^{b}$~Laboratory of Crystallography, University of Bayreuth, D-95440 Bayreuth, Germany.}}
\footnotetext{\textit{$^{c}$~Diamond Light Source Ltd., Harwell Science and Innovation Campus, Didcot,
Oxfordshire, OX11 0DE, U.K. }}

\section{Introduction}

The discovery that many metal--organic frameworks (MOFs) respond mechanically to external stimuli in extreme and counterintuitive ways (\emph{e.g.}\ negative thermal expansion (NTE),\cite{Zhou_2008,Wu_2008} negative compressibility,\cite{Li_2012,Ogborn:2012,Cairns_2015} breathing transitions,\cite{Serre_2007,Ferey:2009} and amorphisation\cite{Chapman_2009,Bennett_2010}) has focussed attention within the field on establishing composition--property relationships. The hope---as in much of MOF science---is to develop design strategies that will allow the targeted synthesis of MOFs with specific and optimised physical properties.\cite{Allendorf:2008,Horcajada:2008} Given the enormous structural and compositional diversity of this family, the task of establishing design rules has focussed on two separate areas: namely, establishing in turn the roles of network \emph{topology/geometry} and network \emph{composition} in governing elastic response.

With regard to the former, it is now well established that certain network connectivities intrinsically favour counterintuitive mechanical responses.\cite{Sarkisov:2014,Collings:2013,Bouessel:2014,Bennett:2015,Ortiz:2013} Perhaps the best known example is the wine-rack topology, which is associated with anomalous elastic behaviour across a large variety of chemically-distinct MOF families.\cite{Nanthamathee:2014,Hunt:2015,Cai:2014,Henke:2013,Zhou:2013,DeVries:2011,Yang:2009} For a given network topology, changes in geometry (\emph{e.g.}, network angles) can invert the type of anisotropy but the basic mechanical response remains the same.\cite{Collings:2014,Henke:2014}

With regard to network composition, variation of the chemical makeup of MOFs with the same topology will usually influence only the magnitude of mechanical response.\cite{Dubbeldam_2007,Bennett:2010,Henke:2013,Li:2014} For instance, nanoindentation studies of the perovskite-structured [(CH$_3$)$_2$NH$_2$][M(HCOO)$_3$] family (M = Mn, Co, Ni, Zn) showed a strong relationship between ligand field stabilisation energy of the transition metal dication and the mechanical stiffness of the corresponding MOF.\cite{Tan:2012} In the related systems [C(NH$_2$)$_3$][Mn(HCOO)$_3$] and [(CH$_2$)$_3$NH$_2$][Mn(HCOO)$_3$], the hydrogen-bonding strength of the extra-framework alkylammonium cation was found to direct the stiffness and flexibility of the two structures.\cite{Li:2014} In all cases, the fundamental mechanism of elastic response is unchanged by chemical substitution, with the degree of flexibility scaling inversely with strength of interaction.\cite{Ogborn:2012} Precisely the same conclusion has been reached in similar studies of inorganic frameworks. For example, an investigation of the thermal expansion response of Prussian Blue analogues and related lanthanide hexacyanocobaltates showed that cation size correlates with the magnitude of NTE.\cite{Chapman:2006b,Adak:2011,Duyker:2013} Likewise, cation size also plays an important role in the magnitudes of thermal expansion behaviour in the family of frameworks related to NaZr$_2$(PO$_4$)$_3$.\cite{Petkov:2003} Perhaps the only exceptions to these general rules are in instances where chemical substitution alters the nature of the dominant chemical interaction within a particular MOF.\cite{Nanthamathee:2014,Millange:2008}

One particularly important way in which dense MOFs can differ from ``conventional'' ceramic frameworks is in their capacity to accommodate extra-framework ions that are molecular rather than monatomic in nature. The added complexity of cation \emph{asphericity} is conceptually related to the symmetry-lowering effect of second-order Jahn-Teller instabilities and may play a key role in the lattice dynamics of high-profile hybrid frameworks such as [CH$_3$NH$_3$]PbI$_3$.\cite{Lee_2015} Yet, to the best of our knowledge, there are no systematic studies of the relationship between counterion shape and mechanical response in MOF-type systems.

\begin{figure}
\centering
\includegraphics{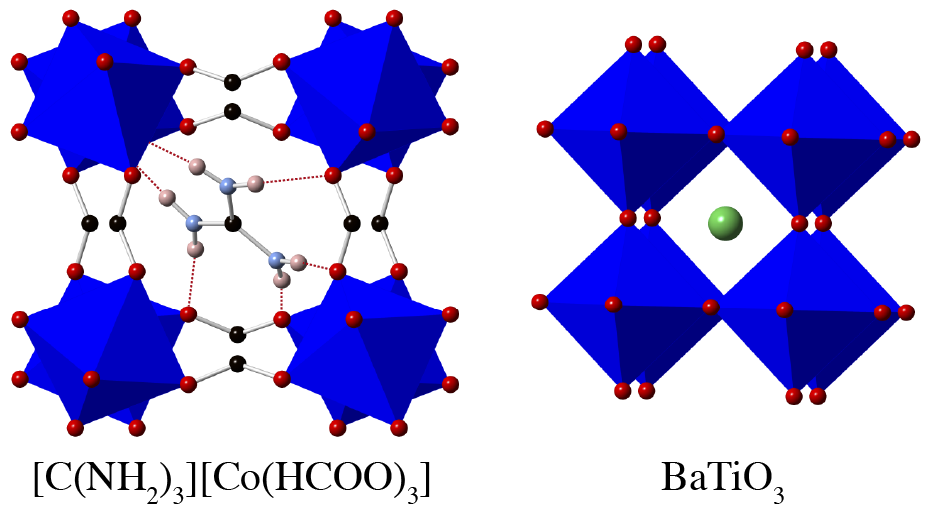}
\caption{ABX$_3$ frameworks of [C(NH$_2$)$_3$][Co(HCOO)$_3$] and BaTiO$_3$.  The hydrogen bonding in [C(NH$_2$)$_3$][Co(HCOO)$_3$] is shown with dotted red lines.  The Co$^{2+}$ and Ti$^{4+}$ coordination environments are represented by polyhedra.}
\label{fig1}
\end{figure}

In this paper, we use thermal expansion measurements to explore the relationship between framework flexibility and chemical composition across 19 MOFs drawn from the widely-studied family of alkylammonium transition-metal (``ABX$_3$'') formates. These compounds have the same general formula [A$^{\textrm{I}}$][M$^{\textrm{II}}$(HCOO)$_3$] and are structurally related to the perovskites [Fig.~\ref{fig1}]. The larger M$\ldots$M separation in formates relative to perovskites ($\sim$6\,\AA\ \emph{vs}.\ 4\,\AA) allows incorporation of molecular cations within the the framework cavities. As in the perovskites there is scope for substitution on both the ``twelve-coordinate'' A-site and on the transition-metal B-site. While the cation charges on each of these sites are fixed at 1+ (A) and 2+ (B) in the formates (unlike perovskites), there is now an additional degree of freedom in terms of the molecular shape at the A-site.

Their relative structural simplicity makes these compounds ideal candidates for a composition/property study such as ours. But the broader family is also of significance from a functional materials viewpoint. ABX$_3$ formates exhibit a variety of useful properties, including ferroelectricity,\cite{Jain:2009, Sanchez-Andujar:2010,Pato-Doldan:2012} ferroelasticity,\cite{Li:2013} and even multiferroic behaviour.\cite{Wang_W:2013,Stroppa:2011,Stroppa:2013,Tian:2014b} Composition/property studies of the ferroelectric response in the A$^{\textrm{I}} = \textrm{N}$H$_4^{+}$ or (CH$_3$)$_2$NH$_2$$^{+}$ members points to the different possible effects of substitution on the alkylammonium site and the transition-metal site.\cite{Xu:2010,Xu:2011,Sanchez-Andujar:2010} Substitution at the A-site allows tuning of the ferroelectric polarisation,\cite{DiSante:2013} whereas variation in transition-metal affects the ferroelectric transition temperature $T_{\rm c}$.\cite{Pato-Doldan:2012,Shang:2014} In this context, framework flexibility may have important implications for A-site orientational ordering and/or mechanisms of accommodating the large strains induced during ferroelectric/paraelectric switching. So this study has the additional relevance beyond establishing composition--property relationships of exploring the role of flexibility in ferroelectric MOFs.

Our paper is arranged as follows. We begin by summarising the synthesis and characterisation techniques used in our study. In the results section that follows, we report the thermal expansion properties of the various ABX$_3$ formates we study. Using the mechanical building unit (XBU) abstraction developed in Ref.~\citenum{Ogborn:2012}, we reduce the experimental lattice parameter data we measure to two characteristic values for each system: namely, the expansivities of the framework struts and of the intra-framework angles. By comparing the magnitudes of these two parameters as a function of framework composition, we are able to establish composition/flexibility relationships for this family.

\section{Experimental Methods}
\subsection{Sample preparation}

\begin{figure}
\centering
\includegraphics{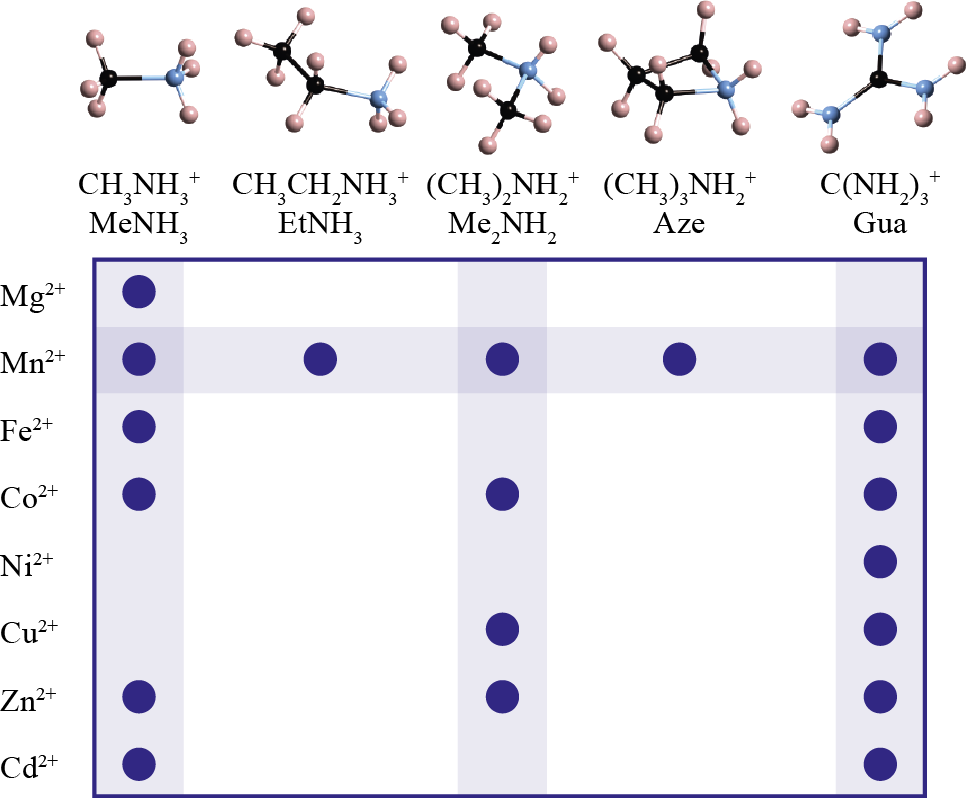}
\caption{Compositions of the various ABX$_3$ formates used in our study. For the structural representations of the different A-site cations given at the top of the table, N atoms are shown in blue, C atoms in black, and H atoms in pink. The shaded regions of the table show the four compositional studies possible in which one or other of the A and B components is kept constant.}
\label{A-cations}
\end{figure}

The 19 structures we investigated share the composition [A$^{\textrm{I}}$][M$^{\textrm{II}}$(HCOO)$_3$], where A$^{\textrm{I}} = \textrm{C}$H$_3$NH$_3$, (CH$_3$)$_2$NH$_2$, CH$_3$CH$_2$NH$_3$, (CH$_2$)$_3$NH$_2$, C(NH$_2$)$_3$ and M$^{\textrm{II}} = \textrm{M}$g, Mn, Fe, Co, Ni, Cu, Zn, Cd [Fig.~\ref{A-cations}].  For simplicity, the [CH$_3$NH$_3$][M(HCOO)$_3$] structures will be referred to as MeNH$_3$--M, where M is the metal cation. Likewise we refer to [CH$_3$CH$_2$NH$_3$][M(HCOO)$_3$] as EtNH$_3$--M, [(CH$_3$)$_2$NH$_2$][M(HCOO)$_3$] as Me$_2$NH$_2$--M, [(CH$_2$)$_3$NH$_2$][M(HCOO)$_3$] as Aze--M (\emph{i.e.}, Aze = azetidinium), and [C(NH$_2$)$_3$][M(HCOO)$_3$] as Gua--M (\emph{i.e.}, Gua = guanidinium).

Our general strategy for preparing [A][M(HCOO)$_3$] samples was as follows.  Methanolic solutions of HCOOH (0.5\,M, 5\,mL) and of the (usually) neutral A-site amine (0.5\,M, 5\,mL) were mixed at the bottom of a glass vial. Onto this solution methanol (2\,mL) was carefully added, followed by a methanolic solution of the transition metal nitrate (0.1\,M, 8\,mL).\cite{Wang_Z:2004} The tube was then sealed and kept undisturbed.  Following precipitation of the product (often as single crystals), the solution was filtered off, and the sample was washed with methanol, dried in air, and ground. The precise species used in the methanolic amine preparation were: methylamine (Acros Organics, 2\,M solution in methanol), dimethylamine (Sigma Aldrich, 2\,M solution in methanol), ethylamine (Sigma Aldrich, 2\,M solution in methanol), azetidine (Sigma Aldrich, 98\%), and guanidine carbonate (Aldrich, 99\%).

\subsection{X-ray powder diffraction}

Synchrotron X-ray powder diffraction data were collected using the I11 beamline\cite{Thompson:2009} at the Diamond Light Source ($\lambda = 0.82715$\,\AA) for each of the 19 different ABX$_3$ compositions given above. Finely-ground powder samples were loaded into 0.5\,mm diameter borosilicate capillaries and mounted on the diffractometer. The Cryostream Plus from Oxford Cryosystems was used to vary the temperature between 110 and 300\,K, and diffraction patterns were collected using the Mythen2 position sensitive detector (PSD). For each sample, data collection started at 300\,K, before cooling to 110\,K at a rate of 5\,K\,min$^{-1}$. To minimise the effects of beam damage, data were not collected continuously but rather at intervals of 10\,K during this cooling process. Once the minimum temperature was achieved, the goniometer was translated to allow a fresh part of the sample to be illuminated during heating. The sample temperature was then increased to 300\,K at a rate of 6\,K\,min$^{-1}$, with data collected at intervals of 10\,K. For all data collected, diffraction patterns were obtained using two separate measurements (5\,s each) at different angular orientations offset by 0.25$^\circ$ that were subsequently merged.\cite{Thompson:2011}

Structural models were determined from the powder diffraction patterns using
Rietveld refinement as implemented in the {\sc{topas}} software (academic
version 4.1).\cite{Coelho:2007} For most of the systems studied, the
corresponding crystal structures are already well known; the literature
values were used as a starting model for Rietveld
refinement.\cite{Wang_Z:2004,Hu:2009,Jain:2008,Sanchez-Andujar:2010,Jain:2009,Sletten:1973,Kong:2006}
To the best of our knowledge, the structures of MeNH$_3$--M (M = Mg, Fe, Co,
Zn, Cd) have not been determined previously; we found these to be
isostructural to MeNH$_3$--Mn and we used the known structure of this
compound as a starting point for Rietveld refinement. In the case of
Gua--Cd---which is also previously uncharacterised---we made use of both
single-crystal X-ray diffraction$^{\ddag}$ and our synchrotron X-ray powder
diffraction measurements to determine a relevant structural model (see SI for
further details).

Our Rietveld refinements made use of  a number of chemically-informed restraints and constraints. Bonding restraints were applied to the metal--formate bonds, while rigid bodies were used to model the organic molecular units (\emph{i.e.}~the A-site cations and the formate ligands).  The bond lengths and internal angles defining the rigid bodies were refined for the initial temperature point. After refinement of the ambient-temperature structure for all [A][M(HCOO)$_3$] compounds, subsequent diffraction patterns were refined sequentially using the refinement model from the previous temperature as a starting model.  In these sequential refinements, the free variables were: a polynomial background function, the lattice parameters, the atomic coordinates, a scale factor, and peak shape parameters.  The molecular geometries of the formate anions and the A-site cations were modelled as rigid bodies with translational and rotational degrees of freedom. The Rietveld fits to the ambient and 110\,K temperature points are given as SI.

\section{Results and Discussion}

\subsection{Thermal expansivities and XBU analysis}
\begin{figure*}
\centering
\includegraphics{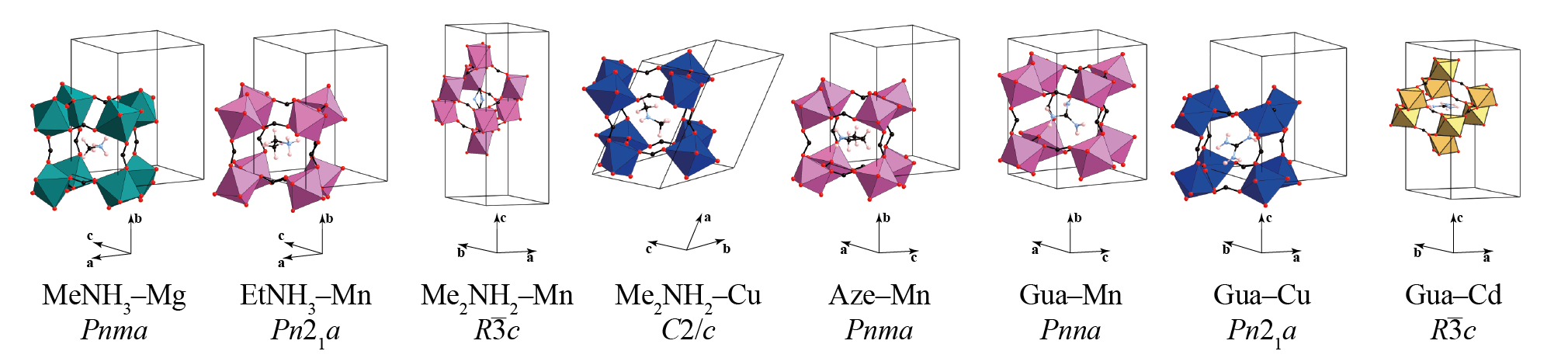}
\caption{Relationship between unit cell and perovskite-like cubic structural unit for each of the crystallographically-distinct phases in our study. In each case, the transition-metal coordination polyhedra are represented as filled octahedra. C atoms are shown in black, N atoms in blue, O atoms in red, and H atoms in pink.}
\label{figy}
\end{figure*}

Our determination of the thermal expansion characteristics of ABX$_3$ formates is based on the interpretation of temperature-dependent lattice parameters extracted \emph{via} Rietveld refinement of synchrotron X-ray powder diffraction data. While the various phases we study are isostructural in the sense that their topologies are identical, they adopt a variety of different space group symmetries. The various crystal phases included in our study are shown in Fig.~\ref{figy}. What is immediately clear is that direct comparison of lattice expansivities from one system to another is not physically meaningful, since the various structures involve different relationships between the perovskite lattice and the unit cell geometry.

Our solution to this problem is to interpret the lattice parameter changes for a given system in terms of two fundamental XBUs:\cite{Ogborn:2012} the framework strut length $r$ and the intra-framework framework angle $\theta$. For systems of sufficiently high symmetry (\emph{e.g.}\ the rhombohedral structures of Me$_2$NH$_2$--Mn/Co/Zn and Gua--Cd) there is a one-to-one mapping between the lattice parameters and XBU coordinates $r,\theta$. In other words, the elastic behaviour of the framework is completely described by the propensity for hingeing (changes in $\theta$) and network deformation (changes in $r$). For lower-symmetry structures, there will be more than one crystallographically-distinct value of $r$ and/or $\theta$; we use an average value in order to facilitate comparison between different systems (see SI for further details). This means that the relationship between lattice parameters and $r,\theta$ is approximate rather than exact, but we find that in practice the degree of approximation is $\lesssim10\%$, which is sufficient for our analysis.

\begin{figure}[h!]
\centering
\includegraphics{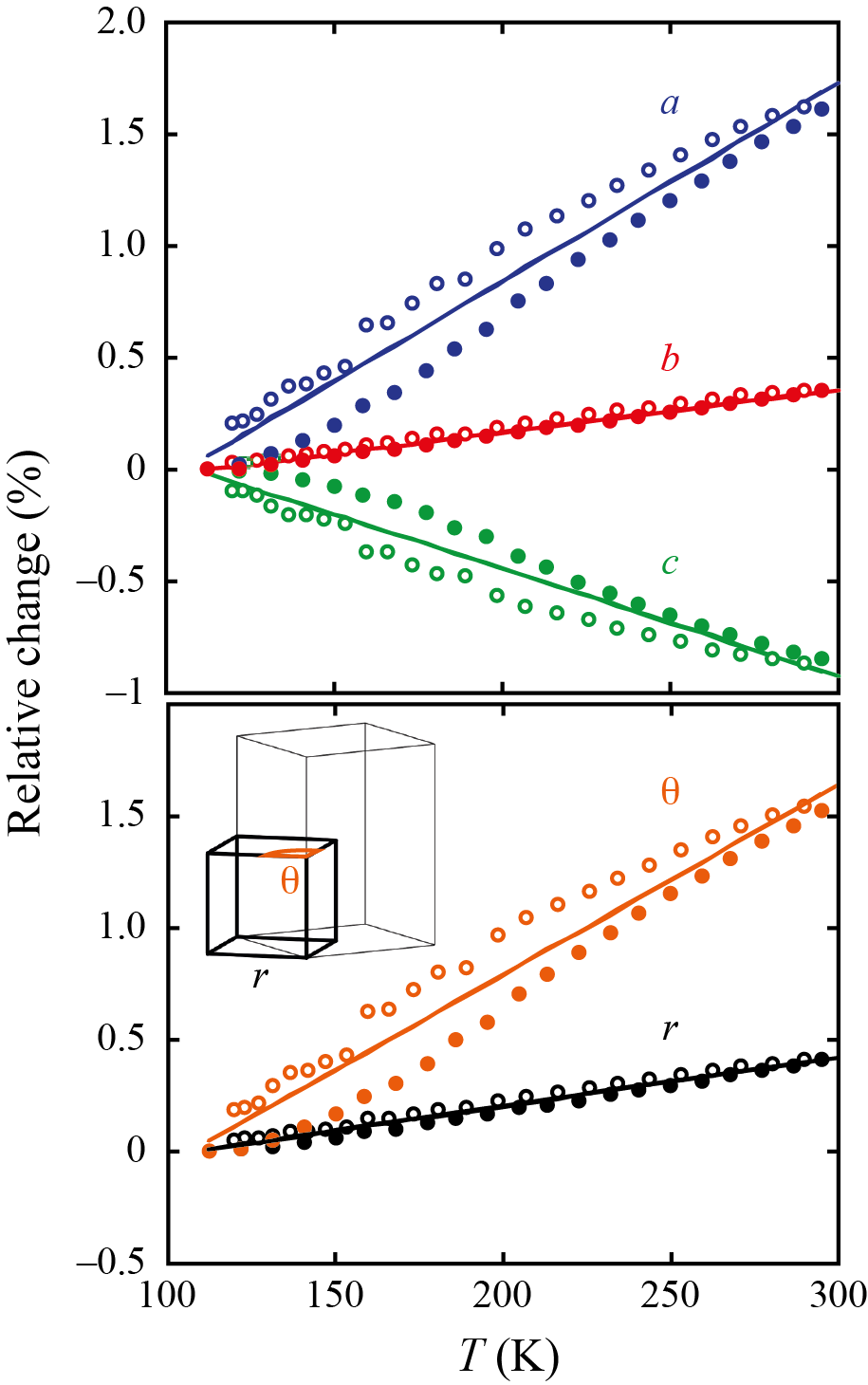}
\caption{Thermal expansion behaviour of MeNH$_3$--Mn, represented in terms of the temperature dependence of lattice parameters (top panel) and XBUs (bottom panel). Data collected during cooling are shown as open circles; those collected during heating are shown as filled circles. The discrepancy between cooling and heating runs likely reflects a combination of beam damage and thermal lag.}
\label{fig34}
\end{figure}

We demonstrate this approach using MeNH$_3$--Mn as a representative example. This compound has the $Pnma$ structure shown in the leftmost panel of Fig.~\ref{figy}. The relative changes in lattice parameters determined using Rietveld refinement against our experimental X-ray powder diffraction are shown in Fig.~\ref{fig34}. These reveal that, on heating, the framework expands rapidly along the $a$-axis and contracts almost as rapidly along the $c$-axis; this collective behaviour corresponds to hingeing of the framework as discussed elsewhere.\cite{Li_2012} The $b$-axis length is unaffected by framework hingeing, and its modest expansion with temperature reflects the intrinsic positive thermal expansion characteristic of M--formate--M linkages.\cite{Ogborn:2012} We quantify the magnitude of thermal response in terms of the coefficients of thermal expansion
\begin{equation}
\alpha=\frac{1}{\ell}\frac{{\rm d}\ell}{{\rm d}T},
\end{equation}
to give $\alpha_a=+88(3)$\,MK$^{-1}$, $\alpha_b=+19.5(4)$\,MK$^{-1}$ and $\alpha_c=-49(2)$\,MK$^{-1}$.\cite{Cliffe:2012}

 \begin{table*}[tb]
\centering
\caption{Lattice and XBU coefficients of thermal expansion for the 110--300\,K temperature range, given in units of MK$^{-1}$. For orthorhombic and hexagonal structures $\alpha_1 = \alpha_a$, $\alpha_2 = \alpha_b$, $\alpha_3 = \alpha_c$; for monoclinic structures the $\alpha_i$ index represents the principal coefficients of thermal expansion.\cite{Cliffe:2012} The expansivities for Aze--Mn given here were determined for the orthorhombic cell characteristic of the ambient-temperature phase; see SI for the principal-axis expansivities of the low-temperature monoclinic phase. $^a$ For this phase, values of $\alpha$ were calculated using cooling data only.}
\label{expansivities_Mformates}
\begin{tabular}{c|c|c|ccc|cc}
A$^{+}$ & M$^{2+}$ & Space group & $\alpha_1$ &  $\alpha_2$ &  $\alpha_3$ & $\alpha_r$ & $\alpha_\theta$  \\
\hline \hline
MeNH$_3$ & Mg  & $Pnma$ & 54.5(1.3)   & 22.1(5)  & $-$20.0(5)   & 20.1(5)  & 46.0(1.1)  \\
                          & Mn  & $Pnma$ & 88(3)     & 19.5(4)   & $-$49(2)    & 21.8(6)  &  84(3)    \\
                          & Fe    & $Pnma$ & 74(2)     & 14.7(4)   & $-$25.1(1.2) & 22.5(6)  &  61(2)   \\
                          & Co   & $Pnma$ & 68.7(1.3)  & 22.3(4)   & $-$28.8(6)  & 21.6(4)  & 61.1(1.2) \\
                          & Zn   & $Pnma$ & 69(3)     & 18.9(9)   & $-$34.6(1.0) & 18.9(1.2) & 65(3)     \\
                          & Cd   & $Pnma$ & 102(7)    & 15.0(1.4)  & $-$61(7)    & 21.5(7)  & 100(8)   \\
\hline
EtNH$_3$& Mn$^{a}$  & $Pn2_1a$ &   44.2(4)  & 33.2(4)  & 4.14(13) & 28.4(3) & 23.88(15) \\
\hline
Me$_2$NH$_2$ & Mn  & $R\bar3c$ & 4.5(5)        & 4.5(5)   & 60.4(1.3)  & 25.9(6) & 24.2(6)   \\
                              & Co   & $R\bar3c$ &  10.6(2)      & 10.6(2)  & 46.5(1.0)  & 24.2(5) & 15.5(3)   \\
                               & Cu                 & $C2/c$ &  $-$14.3(1.0)  & 45.7(1.7)& 57(3)      &   31.7(1.6)      &  24.7(6)         \\
                              & Zn  & $R\bar3c$ &   8.3(9)       & 8.3(9)   & 36.4(4)   & 18.9(1.9)& 12.2(1.2)   \\
\hline
Aze & Mn   & $Pnma$ &   104(8) & 15.3(1.2) &  $-$21(6) & 31.6(1.1)  & $-$78(9) \\
\hline
Gua & Mn      & $Pnna$ & 42.2(6)    & 30.0(4)   & $-$10.6(3)  & 19.2(3)   &  $-$32.2(5)   \\
                     & Fe     & $Pnna$ & 33.1(4)   & 31.6(3)    & $-$1.5(2)   & 20.2(2)   & $-$21.2(3)  \\
                     & Co    & $Pnna$ & 36.6(6)    & 25.8(5)   & $-$6.7(2)    & 17.5(3)   &  $-$26.3(4) \\
                     & Ni   & $Pnna$ & 27.1(8)     & 22.4(6)   & 3.59(16)    & 17.1(5)   & $-$14.3(4)  \\
                     & Cu   & $Pn2_1a$ & 45.9(1.8)    & 21.4(9)   & 1.4(8)      & 22.0(1.1)   & $-$27.2(7)  \\
                     & Zn     &$Pnna$ &  30.7(1.9)    & 21.9(9)   & $-$5.3(6)   & 14.9(7)   & $-$21.9(1.4)\\
      & Cd   & $R\bar3c$ & $-$16.8(9) & $-$16.8(9) & 106(3)      & 16.3(3)   &  $-$43.7(1.3)\\  \hline
\end{tabular}
\end{table*}

For this particular structure, the projection from lattice parameter coordinates onto XBUs is given by the pair of equations
\begin{eqnarray}
r&=&\frac{1}{3}\left(\frac{b}{2}+\sqrt{a^2+c^2}\right),\label{xbu1}\\
\theta&=&2\tan^{-1}\left(\frac{a}{c}\right).\label{xbu2}
\end{eqnarray}
The corresponding change in XBU magnitude with temperature is shown in the bottom panel of Fig.~\ref{fig34}, from which it is clear that the thermomechanical response is dominated by hingeing (changes in $\theta$) rather than network deformation. The XBU coefficients of thermal expansion are $\alpha_r=+21.8(6)$\,MK$^{-1}$ and $\alpha_\theta=+84(3)$\,MK$^{-1}$. XBU projection equations for the various crystal structures of Fig.~\ref{fig1} are given as SI.

The extent of approximation in reducing the three lattice degrees of freedom to the two XBU degrees of freedom can be assessed by calculating values of $\alpha$ for the XBU-derived lattice parameters
\begin{eqnarray}
a_{\rm{XBU}}&=&\frac{2r\tan(\theta/2)}{\sqrt{1+\tan^2(\theta/2)}}\\
b_{\rm{XBU}}&=&2r\\
c_{\rm{XBU}}&=&\frac{2r}{\sqrt{1+\tan^2(\theta/2)}}
\end{eqnarray}
obtained \emph{via} inversion of Eqs.~\eqref{xbu1} and \eqref{xbu2}: we find $\alpha_{a_{\rm{XBU}}}=+87(3)$\,MK$^{-1}$, $\alpha_{b_{\rm{XBU}}}=+21.8(6)$\,MK$^{-1}$ and $\alpha_{c_{\rm{XBU}}}=-50(2)$\,MK$^{-1}$. These are all equal within error to the original lattice coefficients of thermal expansion.

The ABX$_3$ lattice and XBU expansivities determined from our entire ensemble of X-ray diffraction data and corresponding Rietveld fits are summarised in Table~\ref{expansivities_Mformates}; the raw lattice parameter data from which these values were derived are given as SI. While the magnitudes of the lattice expansivities vary over two orders of magnitude, we find essentially the same basic thermomechanical response for all ABX$_3$ structures. For example, nearly all of the orthorhombic structures exhibit large positive thermal expansion (PTE) along the $a$-axis (27--104\,MK$^{-1}$), moderate PTE along the $b$-axis (14.7--33\,MK$^{-1}$), and NTE along the $c$-axis ($-$61 to $-$1.5\,MK$^{-1}$).

\subsection{Effect of variation in M$^{2+}$}

\begin{figure*}
\centering
\includegraphics{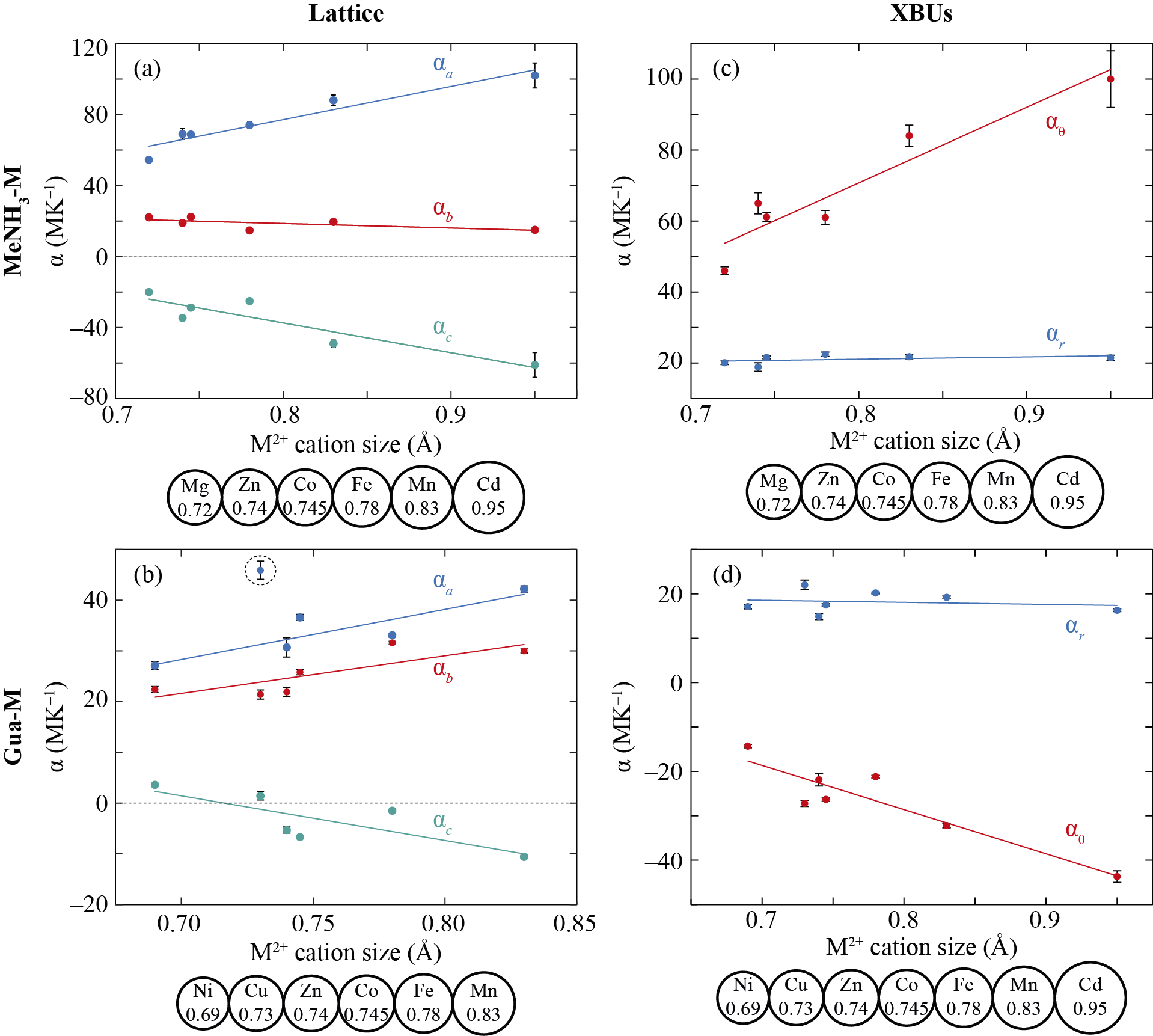}
\caption{Lattice expansivities as a function of metal cation size for (a) MeNH$_3$--M and (b) Gua--M families (only data for the orthorhombic members are shown).  The dotted ellipse in (b) indicates the $\alpha_a$ value for Gua--Cu, which is not included in the linear fit shown here as a guide-to-the-eye. XBU expansivities calculated for complete (c) MeNH$_3$--M and (d) Gua--M families. Ionic radii are taken from Ref.~\citenum{Shannon:1976} for octahedrally coordinated (high-spin) M$^{2+}$ cations.}
\label{CTE_lp_MeNH3_Gua}
\end{figure*}

In order to investigate the effect of M$^{2+}$ variation on the mechanics of [A][M(HCOO)$_3$] frameworks, we compare our results for the MeNH$_3$--M and Gua--M families---these are the two systems for which we have the greatest diversity of M$^{2+}$ substitution. Fig.~\ref{CTE_lp_MeNH3_Gua} presents graphically the relevant data from Table~\ref{expansivities_Mformates}, organised according to the (Shannon) radii of the M$^{2+}$ ions.\cite{Shannon:1976} Two results are immediately obvious from consideration of the dependence of XBU expansivity on ionic radius. The first is that the value of $\alpha_r$ is essentially composition-independent; the second is that the magnitude of $\alpha_\theta$ increases linearly with increasing M$^{2+}$ radius.

The magnitude of $\alpha_r$ reflects the effect of increased thermal motion on the M$\ldots$M separation across connected M--formate--M links. Our Rietveld refinements do not allow us to comment with certainty on the temperature-dependence of the position, orientation, and/or thermal displacement of the formate anions in the various compounds in the way that has allowed detailed analysis of thermal expansion effects in other dense MOFs.\cite{Collings:2013} Nevertheless a likely origin of the composition-independence of $\alpha_r$ is the balance between two competing effects: on the one hand, larger M$^{2+}$ ions would allow greater vibrational motion of bridging formates (\emph{i.e.}, lower vibrational frequencies); on the other hand, the increased M$\ldots$M separation for larger ions means that formate displacements have less effect on this separation \emph{in relative terms}. So while it is likely that the thermal motion of formate ions increases with increasing M$^{2+}$ radius, the relative effect on $r$ as an XBU remains roughly constant.

By contrast, the value of $\alpha_\theta$ represents the degree of hingeing flexibility in a given framework. The increase in flexibility we observe with increasing cation size could arise from several effects. First, the strength of the metal--formate coordination bond will decrease as larger metal cations are used, which in turn facilitates the movement of this bond. This causes an enhancement of the framework flexibility towards framework hingeing, especially as the neighbouring formate linkers are now further apart within the coordination sphere of the metal cation. In previous compositional dependent studies of Prussian Blue analogues, the strength of the coordination bonds was important for tuning the transverse vibrations of the M--ligand--M strut.\cite{Chapman:2006b} Second, the larger unit cells obtained with larger metal cations may allow greater structural freedom, especially as the hydrogen bonding from the A-site cations to the anionic formates of the framework are at greater distances.  These two factors mean that small metal cations, such as Ni$^{2+}$, will have (\emph{i}) short and strong M--O bonds, restricting M--formate movement, and (\emph{ii}) a small unit cell, thus forming shorter intermolecular contacts between formate linkers, and between the framework itself and the A-site cation; all of these factors will result in a restriction of framework hingeing.

While the values of $\alpha_r$ are essentially identical for both MeNH$_3$--M and Gua--M families ($\sim20$\,MK$^{-1}$), the magnitudes of $\alpha_\theta$ \emph{and their sensitivity to cation radius} are substantially smaller for the latter family than for the former. We discuss this dependence on alkylammonium cation in more detail below, but note here that this difference has the consequence for Gua--M systems with small M$^{2+}$ that $|\alpha_\theta|\sim|\alpha_r|$ and so the NTE effect that would ordinarily arise from framework hingeing can be masked by the network deformation (expansion in $r$). This is why Gua--Ni and Gua--Cu do not exhibit NTE along any crystal axis.

\subsection{Effect of variation in A$^{+}$}

Our results for the MeNH$_3$--M and Gua--M families show clearly that variation in A-site cation can affect the extent of framework hingeing observed. To investigate this in more detail, we consider the family of Mn-containing ABX$_3$ formates with A = CH$_3$NH$_3$$^{+}$, CH$_3$CH$_2$NH$_3$$^{+}$, (CH$_3$)$_2$NH$_2$$^{+}$, (CH$_2$)$_3$NH$_2$$^{+}$, and C(NH$_2$)$_3$$^{+}$. The sizes of the different A-site cations were estimated by fitting the atomic coordinates refined from the powder diffraction data to a shape tensor, $L$, using the program {\sc{crystals}}.\cite{Betteridge:2003} This shape tensor represents an anisotropic ellipsoid that contains all atoms within the A-site cation.\cite{Cooper:2004}  The smallest, median, and largest components of this ellipsoid are shown in Table \ref{Acation_size}. We initially use the maximum effective length of the A-site cation ($L_{\rm max}$) as our metric by which to compare the mechanical properties of our different systems. In a another study, a different method was used to calculate the sizes of the A-site cations mentioned here.\cite{Kieslich:2014} The two approaches are consistent with one another except that the order of the Me$_2$NH$_2$$^+$ and Gua$^+$ cations is reversed (see SI for further discussion).
\begin{table}
\centering
\caption[Effective lengths of the A-site cation molecular ellipsoids]{Computed principal axis effective lengths (given in units of \AA$^2$) for each of the A-site cation centroids in the host manganese formate, calculated within {\sc{crystals}}.\cite{Betteridge:2003,Cooper:2004} The asphericity $b$ (also given in \AA$^2$) is defined by Eq.~\eqref{asphericity}.}
\label{Acation_size}
\begin{tabular}{c|ccc|r}
A$^{+}$& $L_{\textrm{min}}$ & $L_{\textrm{med}}$ & $L_{\textrm{max}}$ & \multicolumn{1}{c}{$b$} \\ \hline\hline
MeNH$_3$ & 0.54 & 0.57 & 1.88  & $-$0.64  \\
EtNH$_3$ & 0.65 & 0.88 & 2.95 & $-$0.92 \\
Me$_2$NH$_2$ & 0.71  & 1.21  & 3.25  & $-$0.77  \\
Aze & 0.82  & 1.92  & 1.98 & 0.52 \\
Gua & 0.48 & 2.39 & 2.52  & 0.89  \\\hline
\end{tabular}
\end{table}


\begin{figure}
\centering
\includegraphics{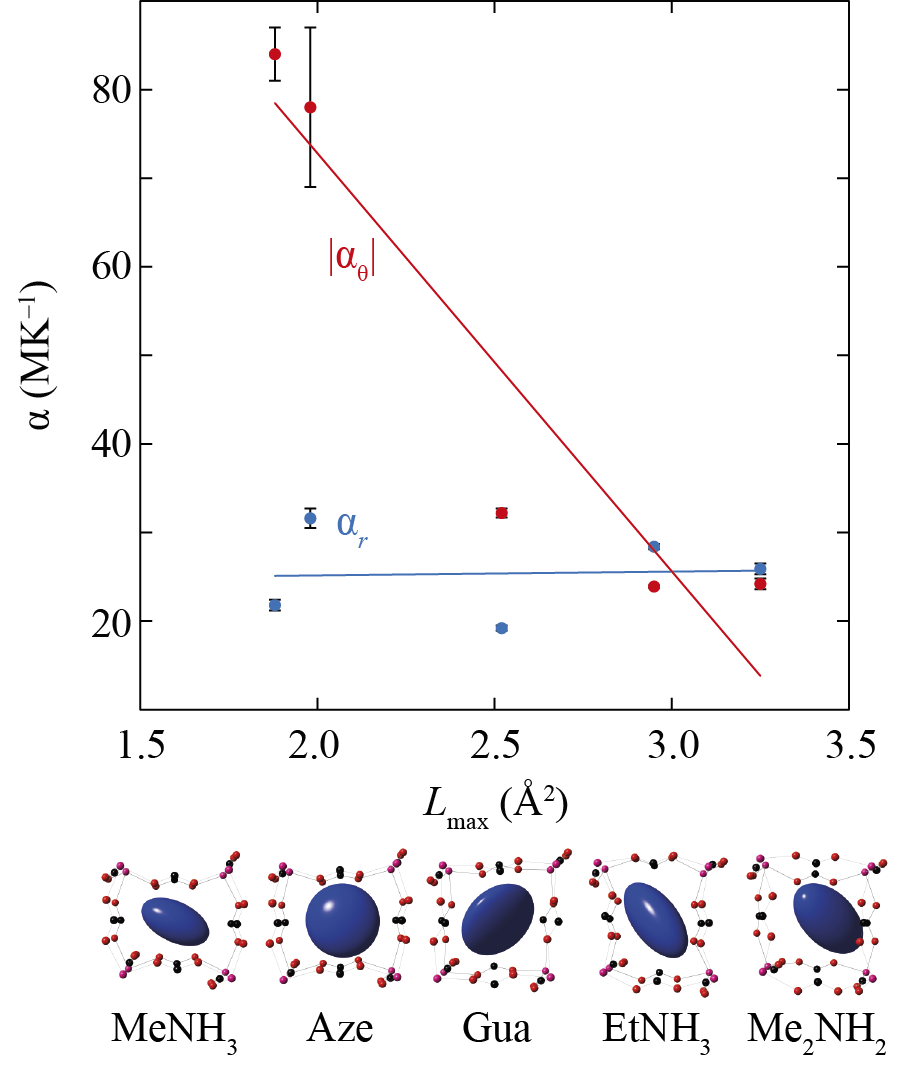}
\caption{XBU expansivities determined for MeNH$_3$--Mn, Aze--Mn, Gua-Mn, EtNH$_3$--Mn, and Me$_2$NH$_2$--Mn, given as a function of maximum effective A-site cation length.  The cation shape tensor generated using {\sc{crystals}} is represented for each compound below the graph.\cite{Betteridge:2003,Cooper:2004} The expansivities for Aze--Mn have an anomalously large uncertainty as a result of a low-temperature phase transition (see SI for further discussion).}
\label{fig5}
\end{figure}

Figure~\ref{fig5} shows the XBU expansivities for the five Mn-containing formate frameworks, arranged according to the value of $L_{\rm max}$ for their A-site cation. As observed in the previous section, there is essentially no meaningful compositional dependence of $\alpha_r$. By contrast, the magnitude of $\alpha_\theta$ is much reduced for those systems with larger A-site cations. Our use of the absolute value of $\alpha_\theta$ reflects the fact that its sign is not conserved amongst the various compounds we study, a point we will expand upon below. So, at face value our results suggest that bulkier A-site cations inhibit framework flexibility. This may be due simply to steric interactions between the A-site cation and the host framework, with bulkier cations preventing large changes in framework angles. From a chemical viewpoint it is likely that the different H-bonding strengths of the various A-site cations will also impact the framework flexibility;\cite{Li:2014} however, our data do not allow us to probe changes in hydrogen-bond characteristics with any certainty. Even in these simplistic terms, and taken together with the results of the previous section, our analysis already suggests that the most flexible formate frameworks will be those with large M$^{2+}$ but small A$^{+}$---precisely the combination that (rightly) identifies MeNH$_3$--Cd as the most flexible of the systems we study here [Table~\ref{expansivities_Mformates}].

\begin{figure}
\centering
\includegraphics{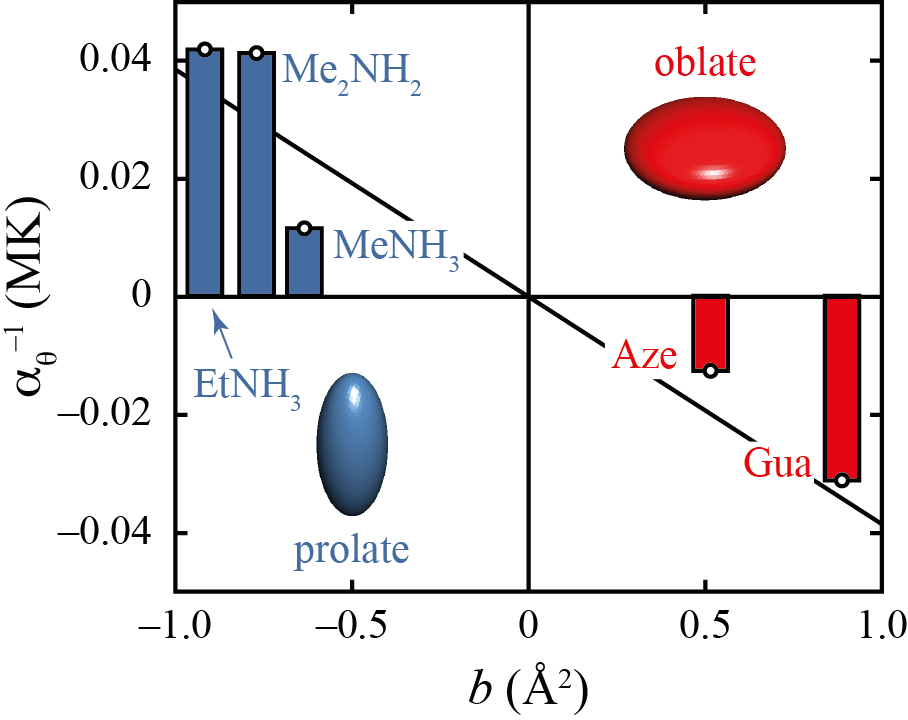}
\caption{Relationship between asphericity $b$ and inverse XBU expansivity for the five Mn-containing ABX$_3$ formates in our study. The direction of framework flexing switches for prolate/oblate A-site cations; the extent of flexibility is largest for small, spherical cations ($b\rightarrow0$).}
\label{fig5_alg}
\end{figure}

One of the key motivations for our study was to understand the relationship between A-cation \emph{shape} and framework mechanics. There are a large number of metrics to quantify shape, but we constrain ourselves here (given that we have but five data points to compare) to the straightforward notion of \emph{asphericity}.\cite{Theodorou_1985} The asphericity parameter
\begin{equation}
b=L_{\rm med}-\frac{1}{2}\left(L_{\rm min}+L_{\rm max}\right)\label{asphericity}
\end{equation}
takes into account both the size of an object and the anisotropy of the shape tensor $\mathbf L$. The parameter $b$ assumes negative values for objects with prolate asphericity, positive values for oblate objects, and is zero for isotropic shapes. The various A-site cations we consider span a range of asphericities $-1\lesssim b\lesssim1$\,\AA$^2$ [Table~\ref{Acation_size}]. Re-ordering the hingeing XBU expansivities according to these values, we find a clear distinction between prolate and oblate A-cations, for which $\alpha_\theta>0$ and $\alpha_\theta<0$, respectively.\footnote{Note that in order for the sign of $\alpha_\theta$ to have any physical meaning, we are careful to define the value of $\theta$ in a consistent way for all frameworks we study. In particular, it is given as the obtuse angle in the rhombic face of the perovskite-like cube. Hence $\alpha_\theta>0$ implies that the cube geometry becomes more distorted with increasing temperature, and $\alpha_\theta<0$ reflects a tendency to adopt an increasingly cubic geometry on heating.} This switching in sign of anisotropic mechanical response as a consequence of cation asphericity is strongly reminiscent of the critical ratios that demarcate linear and area negative thermal expansion in a range of molecular framework materials.\cite{Collings:2014} As in those systems, the magnitude of hingeing response is actually larger for systems closer to the critical geometry (here, $b=0$). Our data show an approximately-linear relationship between $b$ and the inverse expansivity $\alpha_\theta^{-1}$ [Fig.~\ref{fig5_alg}].

So, from a design perspective, our analysis suggests that it is not only the A-cation size that affects framework flexibility, but also its anisotropy. Hence we can explain why EtNH$_3$--Mn is less flexible than Aze--Mn, despite the fact that EtNH$_3^+$ is the smaller cation by $\sim$25\% (see SI for details). This relationship also explains why the qualitative ordering of Fig.~\ref{fig5} in terms of $L_{\rm max}$ is physically meaningful.

\section{Concluding Remarks}

The relationships we establish between cation size and shape provide a set of straightforward structure--property rules which can be implemented as part of the strategic design of functional MOFs. The basis of these relationships is the correlation between cation size and M--formate bonding strength, which in turn affects the mechanical properties.   In a computational study of the [Me$_2$NH$_2$][M(HCOO)$_3$] family, it was shown that the calculated Bader charges on the metal cations exhibited the following order: Mn $>$ Fe $>$ Zn $>$ Co $>$ Ni.\cite{Kosa:2014} This order was also suggested to represent the M--formate bonding strength, where Mn--formate exhibits the weakest bond (highest ionic component) and Ni--formate the strongest (greatest covalent component), and furthermore, the ordering is very similar to that of the metal cation size. There is a small discrepancy with the Zn and Co placement between the Bader charge and metal cation size sequences; however, as there is  very little difference in the sizes (0.74 \emph{vs}.~0.745\,\AA), the variation is not significant.  This small discrepancy can be also seen in the coefficients of thermal expansion, which vary subtly in the order for these two metal cations [Fig.~\ref{CTE_lp_MeNH3_Gua}].

In a nanoindentation study of the [Me$_2$NH$_2$][M(HCOO)$_3$] family, it was suggested that the metal ligand field stabilisation energy (LFSE) directs the resulting mechanical properties.\cite{Tan:2012} The LFSE order of Mn $=$ Zn $<$ Co $<$ Ni was reproduced in the magnitudes of the Young's Moduli of these compounds.\cite{Tan:2012} This finding places compounds which contain either Mn$^{2+}$, Zn$^{2+}$, or Cd$^{2+}$ on the same level of structural flexibility, which is not consistent with the thermal expansivity characteristics we measure here. Likewise, a Brillouin scattering study of related [NH$_4$][Mn(HCOO)$_3$] and [NH$_4$][Zn(HCOO)$_3$] structures indicated increased stiffness for M = Zn$^{2+}$ than M = Mn$^{2+}$.\cite{Maczka:2014} The thermal expansion data from this study and others\cite{Chapman:2006b} also show increased stiffness for Zn$^{2+}$-containing compounds compared to Mn$^{2+}$. This discrepancy could be due to the different experimental methods used: in the case of nanoindentation experiments, a uniaxial force is exerted upon the crystal, while upon temperature variation, an isotropic external stimulus is applied. One obvious and straightforward experiment would be to carry out a comparative nanoindentation study of Zn- and Cd-containing formate frameworks, since these species correspond to very different ionic radii but identical LFSEs.

A clear result of our study has been to show that the presence of A-site cations within the pores of the metal formate framework plays an important role in controlling the structural flexibility of the material. In particular, longer A-site cations cause a significant decrease in the framework hingeing observed.  Thus these structures are not expected to give rise to anomalous mechanics, such as NTE. Instead their mechanics will be dominated by the behaviour of the M--linker--M units. In addition, the direction of framework hingeing can be switched by using differently-shaped A-site cations within the framework pores.  In the case of oblate A-site cations, the framework hingeing is directed towards a convergence of framework angles (such as 90$^{\circ}$ for a 2D wine-rack) upon heating; whereas there is a divergence of framework angles in the case where prolate A-site cations are used. The framework hingeing direction could have implications for ferroelectric transitions arising from reorientation of extra-framework molecular ions. Stronger host--guest interactions might be expected for pore shapes which mimic the shape of the molecular ion. Thus ferroelectric ordering may be stabilised in cases where the pores vary (\emph{i.e.}\ framework angle evolve) towards a shape that mimics that of the molecular ion.

In summary, our work has shown that metal cation size correlates well with expansivity magnitudes. For the octahedral cation site, cations with smaller radii give rise to stiffer frameworks, while the larger metal cations are associated with the more extreme framework flexibility. At the A-site, cation size is also found to affect framework flexibility, though in the opposite sense: framework hingeing is constrained when long A-site cations are present within the framework pore. Moreover, the type of asphericity of the A-site cation determines the direction of framework hingeing. This is most likely to be observed only in dense MOFs, where the vibrational motion of the anisotropic A-site cation is of greater importance.  The most readily applicable rule for rational mechanical design of MOFs involves that of the metal cations, where simple evaluation of its size within different oxidation states or coordination environments can provide a scale of flexibility within a series of isostructural frameworks.

\subsection*{Acknowledgments}

The authors thank the EPSRC (grant no.\ EP/G004528/2) and ERC (grant no.\ 279705) for financial support. This work was carried out with the support of the Diamond Light Source.

\renewcommand\refname{}
\section*{Notes and references}

\footnotesize{ $^{\ddag}$  Single-crystal X-ray diffraction data for Gua-Cd
were collected using a Nonius KappaCCD diffractometer or an Oxford
Diffraction (Agilent) SuperNova diffractometer fitted with an Oxford
Cryosystems Cryostream 600 Series/700 Plus open flow nitrogen cooling device.
\textsc{denzo/scalepack}\cite{Otwinowski_1997} or CrysAlisPro were used for
data collection and reduction as appropriate.  In general, the structures
solved \textit{ab initio} using \textsc{sir}92\cite{Altomare_1994} although
coordinates from 150\,K were used as a starting model for the other
temperatures. All structures were refined with full-matrix least-squares on
$F^{2}$ using \textsc{crystals}.\cite{Betteridge:2003,Parois_2015}  Hydrogen
atoms were generally visible in the difference Fourier map and treated in the
usual manner.\cite{Cooper_2010}  Full structural data are included in the
Supplementary Information (SI) and have been submitted to the CCDC as numbers
1420162--1420165. These data can also be obtained free of charge from The
Cambridge Crystallographic Data Centre via
http:$//$www.ccdc.cam.ac.uk$/$data\_request$/$cif.

}

\vspace{-0.8cm}
\balance \footnotesize{
\bibliography{dalton_2015_abx3} 
\bibliographystyle{rsc} 
}

\end{document}